\documentclass[natbib]{svjour3}    
\journalname{Celestial Mechanics and Dynamical Astronomy}

\smartqed  

\usepackage{amsmath,amsfonts,amssymb,amscd,wasysym}
\usepackage{graphicx}
\usepackage{mathptmx}
\usepackage{natbib}
\usepackage{latexsym}
\usepackage{marvosym}

\newcommand{\Z}{\mathbb{Z}}
\newcommand{\N}{\mathbb{N}}

\def\E{\mathcal E}

\def\H{\mathcal H}

\let\e=\epsilon

\newcommand{\be}{\begin{equation}}
\newcommand{\ee}{\end{equation}}
\newcommand{\ba}{\begin{eqnarray}}
\newcommand{\ea}{\end{eqnarray}}
\newcommand\beq[1]{ \begin{equation}\label{#1} }
\newcommand{\eeq}{ \end{equation} }

\newcommand\beqa[1]{ \begin{eqnarray} \label{#1}}
\newcommand{\eeqa}{ \end{eqnarray} }
\newcommand{\beqano}{ \begin{eqnarray*} }
\newcommand{\eeqano}{ \end{eqnarray*} }

\newcommand{\nn}{\nonumber \\}
\def\n{\vskip.2cm \noindent}

\def\bbm[#1]{\mbox{\boldmath $#1$}}

\begin{document}

\title{Normal forms for the Laplace resonance}


\author{Giuseppe Pucacco
}


\institute{              G. Pucacco \at Dipartimento di Fisica and INFN -- Sezione di Roma II,
Universit\`a di Roma ``Tor Vergata", \\
Via della Ricerca Scientifica, 1 - 00133 Roma\\
\email{pucacco@roma2.infn.it} }
\maketitle

\begin{abstract}
We describe a comprehensive model for systems locked in the Laplace resonance. 
The framework is based on the simplest possible dynamical structure provided by the 
Keplerian problem perturbed by the 
resonant coupling truncated at second order in the eccentricities. The reduced Hamiltonian,  
constructed by a transformation to resonant coordinates, 
is then submitted to a suitable ordering of the terms and to the study of its equilibria. 
Henceforth, resonant normal forms are computed.  
The main result is the identification of two different classes of equilibria. In the first 
class, only one kind of stable equilibrium is present: the paradigmatic case is that of the 
Galilean system. In the second class, three kinds of stable equilibria are possible and 
at least one of them is characterised by a high value of the forced eccentricity for the 
`first planet': here the paradigmatic case is the exo-planetary system GJ-876, 
  in which the combination of libration centers admits triple conjunctions 
otherwise not possible in the Galilean case. 
The normal form obtained by averaging with respect to the free eccentricity oscillations, 
describes the libration of the Laplace argument for arbitrary amplitudes and allows us 
to determine the libration width of the resonance. The agreement 
of the analytic predictions with the numerical integration of the toy models is very good. 



\end{abstract}

\section{Introduction}
\label{intro}

Multi-resonant planetary problems are extremely interesting both for theory and applications. The prototypical example is given by the system 
composed of Jupiter and its first three Galilean satellites, Io, Europa and Ganymede. The satellites 
are phase-locked into the so-called \sl Laplace resonance \rm \citep{FeBook,MD}. This implies the
ratio 4:2:1 of the mean motions and a fixed value of the
relative precession of the \sl peri-Joves \rm of Io and Europa. 
Another well known example is GJ-876 \citep{RG876} which is an exo-planetary system with the same multi-resonant structure. 
Other extrasolar systems 
with the same mean motion resonances are under investigation \citep{MO13,PiBaMo} 
in the ever growing census of these systems (see e.g. \cite{FLR}). 
However, multi-resonant systems are not so common \citep{B15} even with more general resonant combinations \citep{GaCoBa}, implying complex
general questions about their origin and stability, in particular in the case of compact systems.

The motion of the Galilean system is characterised by a regular behaviour: out of the four resonant angles combining longitudes and arguments
of peri-Joves, three are \sl librating \rm with small amplitudes and one is \sl rotating \rm \citep{ShMa} with almost constant angular velocity. On the other hand, 
the same resonant angle is reported to have a chaotic evolution in the case of GJ-876 \citep{BDH,BEN}. It is therefore important to understand 
the structure of the regular part of phase-space and the possible origins of chaotic dynamics. Moreover, the long-term evolution of these systems is affected 
by dissipative effects \citep{YoPe,BaMo2,PiBaMo}: actually, in the Galilean system, the coupling between tides and the internal dynamics plays an essential role in the 
approach to the resonance and its subsequent evolution. The fundamental work of \citet{YoPe} introduced an analytical model and subsequent studies \citep{HeTides,Ma,ShMa,LSF}
have extended the treatment with semi-analytical or numerical methods. To have a simple but reliable model for the conservative dynamics may offer clues to investigate also
the cases in which non-conservative effects have different origins like in protoplanetary disks.

The purpose of this work is to elaborate on the following two problems:

1. All analytical and numerical computations so far \citep{Li77,HM81,MuVaMoSc,La-A,La-B} show that the Laplace resonance is quite stable but at the same time very sensitive to the values of orbital elements. The first question is therefore to be able to predict the extent of the resonance domain, that is to devise a reduction process able to identify and compute the width of the resonance in terms of the parameters of the system.

2. The Laplace status of the Galilean satellites is very well understood and described in terms of three combination angles librating around an equilibrium value: however, these equilibrium values (or their symmetric equivalent) are clearly different from those reported for the GJ-876 system \citep{RG876,BEN}. This discrepancy raises issues concerning the stability of this configuration. The second question we want to address refers therefore to the possibility of predicting the location and nature of the equilibria in terms of the parameters of the system, in order to interpret the status of this and other possible configurations.

As said above, there is a fourth combination angle which could also librate but is observed to rotate in the real Galilean system. 
The configuration in which all four combinations librate is known as 
\sl de Sitter equilibrium \rm \citep{Si,BH,ZHB,CePaPu}. Therefore another question related to point 2 above 
is why the observed systems are not in the de Sitter equilibrium and how to predict the 
possible regular/chaotic regimes of the involved arguments. The model we study is strictly related to the classical basic ones, starting from the seminal works of \citet{Si75} and \citet{FeBook} and elaborated mainly by \citet{HeCons,He} and \citet{Ma}. It closely follows the applications done in \citep{CePaPu,CePaPuAA} and generalised in \citep{CePaPuChao}. The substantial difference in the present approach is to modify the choice of variables adapted to the resonance and to convert the system to a slowly perturbed chain of forced harmonic oscillators. 

Actually, the study of mean-motion resonances is one of the pillars of Celestial Mechanics \citep{Poi02,SAObs,W1980,HL} which should now be standard textbook material \citep{MD,MO02,FeBook2007}. However, the intricacies of resonant dynamics often require dedicated expansions and coordinate transformations \citep{HLMM,Moons94,MBFM,BaMo,PiMoCr} for perturbative applications. We have endeavoured to find a framework better suited to first-order multi-resonant planetary problems. The main advantage is to have a suitable action conjugated with the Laplace argument to be used in the construction of a resonant normal form. This leads to a direct evaluation of the libration frequency and of the resonance width. Moreover, with this coordinate choice it is easier to analyse the forced equilibria by finding an efficient strategy to locate additional solutions with respect to the classical ones \citep{Si,Si75}. 

We validate   this approach by applying it \rm to the two reference systems mentioned above, the Galilean satellites and the GJ-876 system. 
They result in two toy models limited to   second-order \rm expansions in the eccentricities so to reduce as much as possible mathematical complexity. 
  The Galilean case is already described quite faithfully at first-order and possible discrepancies between predictions of the model and observations are only quantitative; 
they can be reduced with higher-order expansions. 
In the exo-planetary instance the first-order model is able to predict several peculiarities: in particular, the different nature of the Laplace dynamics 
(when compared to the Galilean system) and the high values of the forced eccentricities. However, the reported libration 
centers are correctly predicted by the model only when including second-order terms in the eccentricities. \rm

The plan of the paper is as follows: in section \ref{sec:model} is presented the basic starting Hamiltonian model; in section \ref{sec:bke} 
 the truncated series of the model is set with a suitable book-keeping order and are determined its equilibrium solutions; in section \ref{sec:nfs} 
 are computed the Laplace normal form and its variants; in section \ref{SAPP} the model is applied to observed systems; in section \ref{SCON} 
 conclusions are drawn.

\section{The simplified starting model}
\label{sec:model}

We consider a 1+3-body self-gravitating system in which three `planets' of mass $m_k, k=1,2,3$ 
are orbiting around a `star' $m_0$ with mean motions close to the ratios $n_1 / n_2 = n_2 / n_3 = 2$. We work in modified Delaunay variables 
$$
L_k, \ \lambda_k, \quad P_k=L_k\left(1-\sqrt{1-e_k^2}\right), \ p_k=-\varpi_k $$ 
and, in the usual case of small eccentricities, almost always we will assume that
$$
P_k \simeq \frac12 L_k e_k^2.$$
Using a Jacobi coordinate system \citep{He}, the Kepler part of the Hamiltonian is given by
\be\label{Hkep}
H_{Kep }=-\frac{Gm_0m_1}{2a_1}-\frac{G(m_0+m_1)m_2}{2a_2}-\frac{G(m_0+m_1+m_2)m_3}{2a_3} = - \frac12 \sum_{k=1}^3 \frac{M_k^2 \mu_k^3}{L_k^2}\ ,
\ee
where $a_k $ are the semi-major axes, the $L_k$ are defined as
\be\label{LP}
L_k=\mu_k\sqrt{M_k a_k}\ 
\ee
and
\beq{mudef}
\mu_k = \frac{M_{k-1} m_k}{M_k}, \quad (k=1,2,3)
\eeq
with $M_0=Gm_0$ and 
$$ M_k = M_0 + G \sum_{j=1}^k m_j. $$
In order to implement normalisation, it is useful to expand the Keplerian 
part as follows \citep{BaMo}
\be\label{Hkep_esp}
H_{Kep \ exp}= \sum_{k=1}^3 \left[\overline  n_k (L_k - \overline L_k) - \frac32 \eta_k (L_k - \overline L_k)^2 \right] \ .
\ee
$\overline L_k$ are three `nominal' values of the first actions
and 
\be\label{neta}
\overline n_k = \sqrt{\frac{M_k}{\overline a_k^3}} = \frac{M_k^2 \mu_k^3}{\overline L_k^3}, \qquad \eta_k = \frac{\overline n_k}{\overline L_k} 
\ee
are evaluated at nominal values. Exactly resonant mean motions such that
\beq{resndef}
\overline n_1 = 2 \overline n_2 = 4 \overline n_3\eeq
would provide the set of resonant nominal actions, but
we remark that the choice of nominal elements is rather arbitrary and we could as well choose not strictly resonant mean motions. 
Pros and cons of these choices have been the subject of detailed analyses in several papers devoted to mean-motion resonances 
for which we refer to \citet{BaMo} and references therein. 

The {  disturbing} function,   as usual in the general case of 
first-order resonances \citep{FeBook,BaMo2,Pa15}, \rm is limited to the first-order terms in the 
expansion in the eccentricities $e_k$.  After averaging with respect to the non-resonant angles we keep the terms in the following sum:
\ba\label{Hsat1}
H_{pert}&=&-{{Gm_1m_2}\over a_2}\left\{{B_0}(\rho_{12})+
f_1(\rho_{12})e_1\cos(2\lambda_2-\lambda_1-\varpi_1)+
f_2(\rho_{12})e_2\cos(2\lambda_2-\lambda_1-\varpi_2)\right\}\ \nonumber\\
&&-{{Gm_2m_3}\over a_3}\left\{{B_0}\left(\rho_{23}\right)+
f_1(\rho_{23})e_2\cos(2\lambda_3-\lambda_2-\varpi_2)+
f_2(\rho_{23})e_3\cos(2\lambda_3-\lambda_2-\varpi_3)\right\}\ \nonumber\\
&&-{{Gm_1m_3}\over a_3}\left\{{B_0}(\rho_{13})\right\} + {\rm O} (e_k^2)\ , \ea
where the coefficients ${B_0}$ and $f_k$ are defined as
\ba\label{gammak}
B_0(\rho)&=&\frac12 {b_{1/2}^{(0)}(\rho)} - 1, \\
f_1(\rho) &=&  {1\over 2}\left(4+\rho{d\over{d\rho}}\right)b_{1/2}^{(2)}(\rho),\\
f_2(\rho) &=& -{1\over 2}\left(3+\rho{d\over{d\rho}}\right)b_{1/2}^{(1)}(\rho)+2\rho
\ea
and the $b_{s}^{(j)}(\rho)$ are the \sl Laplace
coefficients \rm \citep{MD}, with the ratios $\rho_{ik}=\overline  a_i / \overline a_k,
(i,k=1,2,3)$ usually fixed at their nominal values.
We will scale physical units by choosing units of time and length such that $Gm_0=1$ and $a_1=1$ and therefore it is natural to introduce the parameters
\ba
\varepsilon_k &=& \frac{m_k}{m_0} , \quad k=1,2,3, \label{amg} \\
\overline m_A &=& \frac{m_A}{m_1} = \frac{\varepsilon_A}{\varepsilon_1}, \quad A=2,3\label{ams} .
\ea

\subsection{Henrard-Malhotra variables}
We assume that the 
disturbing function is expressed in terms of modified Delaunay variables. With an aim to investigate the Laplace resonance, it is customary to use the following\footnote{The notations are slightly different from the usual ones: in the following we adopt almost systematically the `Capital-lower case' convention for momenta and coordinates.} \sl
Henrard-Malhotra \rm coordinate transformation $(L_k, P_k, \lambda_k, p_k) \longrightarrow (Q_{\alpha}, q_{\alpha}), \alpha =1,\dots,6,$ \citep{He,Ma}, 
\ba\label{cootr}
q_1&=&2\lambda_2-\lambda_1+p_1 \ , \quad\quad Q_1 = P_1 \ ,\label{q1}\\
q_2&=&2\lambda_2-\lambda_1+p_2 \ , \quad\quad Q_2 = P_2 \ , \label{q2}\\
q_3&=&2\lambda_3-\lambda_2+p_3 \ , \quad\quad Q_3 = P_3 \ , \label{q3}\\
q_4&=&3\lambda_2-2\lambda_3-\lambda_1 \ , \quad \;\;
Q_4 = {\scriptstyle{\frac13}}\left(L_2-2(P_1+P_2)+P_3\right) \ , \label{q4}\\
q_5&=&\lambda_1-\lambda_3 \ , \quad\quad\quad\quad \;\;
Q_5 = {\scriptstyle{\frac13}}\left(3L_1+L_2+P_1+P_2+P_3\right) \ , \label{q5}\\
q_6&=&\lambda_3\ , \quad\quad\quad\quad\quad\quad \;\;\,
Q_6 = L_1+L_2+L_3-P_1-P_2-P_3  \label{q6}
\ea
which takes into account the resonant combinations of the angles. The list of the old $L$-actions in terms of the new ones is
\ba\label{invcootr}
L_1&=&Q_5-Q_4-Q_1-Q_2 \ ,\label{L1}\\
L_2&=&3Q_4+2 Q_1+2Q_2-Q_3 \ , \label{L2}\\
L_3&=&Q_6-Q_5-2Q_4+2Q_3 \ . \label{L3}
\ea
With this transformation, the model can be expressed as
\beq{HLQ}
H (Q_a,q_a; Q_5,Q_6)=\sum_{n=0}^2 H_n  (Q_a,q_a), \quad  a=1,...,4 \ ,
\eeq
where it is made clear that now the dependence on the new angles is such that $q_5,q_6$ are cyclic and therefore $Q_5,Q_6$ are integrals of motion ($Q_6$ is the total angular momentum). Among the non-trivial momenta $Q_a, \;  a=1,...,4$, the first three, being equal to the $P_k$, are `small' quantities. From \eqref{q4} it is evident that $Q_4$  is instead of the order of $L_2$. We are therefore led to introduce also nominal values\footnote{Here we can consider `nominal' initial conditions: however, a possible choice (made for example by \citet{He}) is that of considering nominal circular orbits so that $\overline P_k = 0$.}  for the $P_k$, say $\overline P_k$, such that 
\beq{NQ4}
Q_4 = {\scriptstyle{\frac13}}\left(L_2-2(P_1+P_2)+P_3\right) = {\scriptstyle{\frac13}}\left(\overline L_2-2(\overline P_1+\overline P_2)+\overline P_3\right) + \widehat Q_4 \ , 
\eeq
where $ \widehat Q_4$ is `small'. Accordingly, the three terms in the Hamiltonian \eqref{HLQ} can be written as
\beqa{HFS}
H_0^{HM}&=& \kappa_1^{HM} (Q_1+Q_2) + \kappa_3^{HM} Q_3 + (\kappa_1^{HM} - \kappa_3^{HM}) \widehat Q_4 \ ,\label{HMzero}\\
H_1^{HM}&=& -\frac32 \left[\eta_1 \right(Q_1+Q_2+ \widehat Q_4 \left)^2 + \eta_2 \right(2(Q_1+Q_2)- Q_3 + 3 \widehat Q_4 \left)^2 + 4\eta_3 \right(Q_3 - \widehat Q_4 \left)^2\right]\ ,\label{HMuno}\\
H_2^{HM}&=& - \alpha \sqrt{2 Q_1} \cos q_1 - \beta_1 \sqrt{2 Q_2} \cos q_2 - \beta_2 \sqrt{2 Q_2} \cos (q_2 - q_4) - \gamma \sqrt{2 Q_3} \cos q_3 \ .\label{HMdue}
\eeqa
In the linear part, $H_0^{HM}$, the new frequencies
\beqa{NFR}
\kappa_1^{HM} &=& 2 n_2 - n_1 + 3 (\eta_1 + 4 \eta_2) (\overline P_1+\overline P_2) - 6 \eta_3 \overline P_3 \ , \label{OFR1}\\
\kappa_3^{HM} &=& 2 n_3 - n_2 - 6 \eta_2 (\overline P_1+\overline P_2) + 3 (\eta_2 + 4 \eta_3) \overline P_3 \ , \label{OFR2}
\eeqa
appear. 
The quadratic part $H_1^{HM}$ accounts for the non-linear dependence on the actions. In the third term, the angle dependent part $H_2^{HM}$, the nominal values of the $L_k$ are used so that, using (\ref{gammak}--\ref{amg}), the coupling parameters 
\beqa{PAR}
\alpha &=& \frac{\overline m_2^2 \e_2 f_1(\rho_{12})}{\overline L_2^2 \sqrt{\overline L_1}} , \label{PAR1}\\
\beta_1 &=& \frac{\overline m_2^2 \e_2 f_2 (\rho_{12})}{\overline L_2^{5/2}} , \\
\beta_2 &=& \frac{\overline m_2 \overline m_3^2 \epsilon_3 f_1 (\rho_{23})}{\overline L_3^2 \sqrt{\overline L_2}} , \\
\gamma &=& \frac{\overline m_2 \overline m_3^2  \epsilon_3 f_2 (\rho_{23})}{\overline L_3^{5/2}} \label{PAR4},
\eeqa
are introduced. 

\subsection{Modified Henrard-Malhotra variables}
The basic structure of the model is that sketched above: we have a Hamiltonian which is the sum of the Keplerian expansion up to second order and a coupling part depending also on resonant combination angles. The magnitude of these terms is ordered in a natural way and the resonant angles determine the form of the canonical transformation to variables adapted to the resonance. However, the transformation introduced in \cite{He} is not unique: in the paper on his version of the model on the tidal evolution of the Galilean satellites, \citet{HeTides} himself uses a different form of the action conjugated to new cyclic angles and therefore to different choices of the conserved actions. For our model, we will instead employ a \sl modified \rm Henrard-Malhotra (MHM) coordinate system in which the momentum $Q_4$, whose dynamical meaning is a little obscure, is replaced by
\beq{Nq4}
\Lambda={\scriptstyle{\frac13}}\left(2 L_1 + L_2- L_3\right),\eeq
which is more naturally associated with the conjugate angle $\lambda=q_4$. Using for simplicity the same notation of the previous section for the unchanged variables, the MHM coordinate set is given by 
\ba\label{Ncoolist}
q_1, q_2, q_3^M, \quad && \quad Q_1,Q_2,Q_3, \nn
\lambda, \quad && \quad  \Lambda, \nn
q_5^M, q_6^M,\quad &&   \quad
 Q_5,Q_6, \nonumber
\ea 
where   the new angles are \rm
\ba\label{Ncootr}
q_3^M&=&2\lambda_2-\lambda_1+p_3 \ , \\
q_5^M&=&\lambda_3-3\lambda_2+2\lambda_1 \ , \\
q_6^M&=&\lambda_2 + {\scriptstyle{\frac13}}\left(\lambda_3 - \lambda_1 \right) \ .
\ea
Now we have
\ba\label{Nq3a}
q_3^M = q_3 + q_4
\ea
and the angles $q_5^M,q_6^M$ are again cyclic with the same associated integrals of motion $Q_5,Q_6$ as before in the original transformation. In the MHM coordinate set, the list of the old $L$-actions in terms of the new ones is
\ba\label{invcootrm}
L_1&=&2Q_5-{\scriptstyle{\frac13}}Q_6 - \Lambda -Q_1-Q_2-Q_3 \ ,\label{L1M}\\
L_2&=&2 (Q_1+ Q_2 + Q_3) + 3 \Lambda + Q_6-3Q_5 \ , \label{L2M}\\
L_3&=&Q_5+{\scriptstyle{\frac13}}Q_6 - 2\Lambda \ . \label{L3M}
\ea
By using the MHM set and introducing the `angular momentum deficit' \citep{LaskarAAL,LaskarPRL,LaskarP}
\beq{amddef}
\Gamma=Q_1+Q_2 + Q_3,
\eeq 
the three terms in the Hamiltonian \eqref{HLQ} now are  
\beqa{HFSM}
H_0^M&=& \kappa_1^M \Gamma + \kappa_4^M \Lambda \ ,\label{HMzeroM}\\
H_1^M&=& -\frac32 \left[(\eta_1 + 4\eta_2) \Gamma^2 + 2(\eta_1 + 6 \eta_2) \Gamma \Lambda + (\eta_1 + 9\eta_2 +4 \eta_3) \Lambda^2\right],\label{HMunoM}\\
H_2^M&=& - \alpha \sqrt{2 Q_1} \cos q_1 - \beta_1 \sqrt{2 Q_2} \cos q_2 - \beta_2 \sqrt{2 Q_2} \cos (q_2 - q_4) - \gamma \sqrt{2 Q_3} \cos q_3.\label{HMdueM}
\eeqa
In the resonant part, with a slight abuse of notation, we have left the standard `third' combination angle $q_3$ in place of $q_3^M - q_4$. In the linear part, $H_0^M$, the new frequencies
\beqa{NFRM}
\kappa_1^M &=& 2 n_2 - n_1 -3 \eta_1 \overline L_1 + 6 \eta_2  \overline L_2 +6(\eta_1 + 3 \eta_2) Q_5- (\eta_1 + 6 \eta_2) Q_6 \ , \label{MFR1} \\
\kappa_4^M &=& 3 n_2 -2 n_3 - n_1 - 3 \eta_1 \overline L_1 + 9 \eta_2  \overline L_2 -6 \eta_3 \overline L_3 \nn 
&& + 3(2 \eta_1 + 9 \eta_2 + 2 \eta_3) Q_5 
- (\eta_1 + 9 \eta_2 - 2 \eta_3)  Q_6\ , \label{MFR2}
\eeqa
appear. These new frequencies seems odd if compared with (\ref{OFR1}-\ref{OFR2}), in which the resonant combinations of mean motions are varied exclusively in terms of the small quantities $\overline P_k$. $\kappa_1^M,\kappa_4^M$ rather depend on the `large' quantities $ \overline L_k, Q_5, Q_6 $. This apparent shortcoming is due to the fact that, in the modified transformation, we did not impose the condition that the fourth momentum is a small quantity obtained by a variation analogous to the introduction of $ \widehat Q_4$. The advantage of this choice roots into the special dynamical role played by the new momentum as will appear clear in the following. Therefore, we keep the definition of these frequencies by taking into account that the issue refers to the role of $Q_5, Q_6 $ themselves. Being them integrals of motion, they can be considered as arbitrary parameters of the model. However, the initial conditions of a given solution fix the value of the integrals of motion and we are interested essentially in initial conditions which are sufficiently close to the resonance. A simple choice could be that of fixing the integrals at nominal values of the elements. However, in the implementation of the model this turns out to be over-restrictive, 
  since it places the unperturbed system at exact resonance producing, as we see in the following, the dynamical instability of the perturbed one. \rm In order to be able to explore the proximity to the resonance, we have found that the best choice is that of fixing nominal resonant values for the mean motions used in the Keplerian expansion, but to leave free the values of $Q_5, Q_6 $ as they are fixed by realistic initial conditions. In this way, (\ref{MFR1}-\ref{MFR2}) take the simplified form
\beqa{NMFRM}
\kappa_1^M &=& 6(\eta_1 + 3 \eta_2) Q_5- (\eta_1 + 6 \eta_2) Q_6 \ , \label{MSFR1} \\
\kappa_4^M &=& 3(2 \eta_1 + 9 \eta_2 + 2 \eta_3) Q_5 
- (\eta_1 + 9 \eta_2 - 2 \eta_3)  Q_6. \label{MSFR2}
\eeqa

\subsection{Poincar\'e variables}
By using Poincar\'e variables $x_k, y_k \; (k=1,2,3)$ defined as
\beqa{}
x_1&=&\sqrt{2 Q_1} \cos q_1, \nn 
y_1&=&\sqrt{2 Q_1} \sin q_1 , \nn 
x_2&=&\sqrt{2 Q_2} \cos q_2, \nn 
y_2&=&\sqrt{2 Q_2} \sin q_2 , \nn 
x_3&=&\sqrt{2 Q_3} \cos q_3^M =  \sqrt{2 Q_3} \cos (q_3 - q_4), \\
y_3&=&\sqrt{2 Q_3} \sin  q_3^M =  \sqrt{2 Q_3} \sin (q_3 - q_4), 
\eeqa
  the angular momentum deficit can be written as \rm
$$ \Gamma (x,y) = \frac12 \sum_{k=1}^3 \left( x_k^2 + y_k^2 \right). $$
Hamiltonian \eqref{HLQ} is then changed into
\beq{HLP}
H_P (x_j, y_j,\Lambda,\lambda)=\sum_{j=0}^2 H_j^P , \quad  k=1,2,3,
\eeq
where, for the MHM coordinate set, we get
\beqa{HFSP}
H_0^{P}&=& \kappa_1^M \Gamma (x,y) + \kappa_4^M \Lambda ,\label{HMzeroMP}\\
H_1^{P}&=& -\frac32 A \Gamma (x,y)^2 - 3 B \Gamma (x,y) \Lambda  -\frac32 C \Lambda^2 ,\label{HMunoMP}\\
H_2^{P}&=& - \alpha x_1 - \beta_1 x_2 - \beta_2 (x_2 \cos \lambda + y_2 \sin \lambda) - \gamma (x_3 \cos \lambda + y_3 \sin \lambda) \ \label{HMdueMP}
\eeqa
%
and we have introduced the shorthand symbols:
\ba
A&=& \eta_1 + 4\eta_2, \label{Adef} \\
B&=& \eta_1 + 6\eta_2,\\
C&=& \eta_1 + 9\eta_2 + 4\eta_3. \label{Cdef} \ea

\section{Book-keeping order of the modified model and its equilibria}
\label{sec:bke}
Hamiltonian \eqref{HLP} has a nice `perturbed oscillators' form which promises to be quite simple to use. 
This setting is reminiscent of that proposed by \cite{Sag} and \cite{Br}. However, in order to proceed with the 
analysis of the dynamics it is useful to perform an appropriate book-keeping of the various terms. 
  At the basis of every perturbative approach, there is a decision about the relative weight of small terms 
defining the perturbation. 
Here there are at least two sources of `smallness': the ratio of masses of the orbiting bodies over that of the central 
body and the eccentricities. The model can be considered of first-order with respect to both: therefore, in order 
to simplify things, a unique order parameter is used, making care of pointing out possible cases in which this 
is no more consistent. \rm

\subsection{Book-keeping order}

We see that, in principle, the following hierarchy   
can be established among variables and control parameters (from now on, to lighten notation, we suppress the $M$ index,   except \rm for the angle $q_3^M$ 
which is important to distinguish from the original $q_3$):

\begin{itemize}
\item Zero-Order quantities: $\Lambda; \kappa_1, \kappa_4, \eta_k, k=1,2,3;$
\item First-Order quantities: $x_k, y_k; \alpha, \beta_1, \beta_2, \gamma;$
\item Second-Order quantities: $Q_k = \left( x_k^2 + y_k^2 \right)/2.$
\end{itemize}
The order can be represented by the power of a book-keeping parameter, say $\sigma$: so a term of Nth-Order is multiplied by the label $\sigma^N$. 
In view of this ordering, we rearrange the model Hamiltonian in the following form:
\beq{HB}
\H (x_k, y_k,\Lambda,\lambda)=\sum_{n=0}^2 \sigma^{2n} \H_n , \quad  k=1,2,3,
\eeq
with
\beqa{HBSM}
\H_0&=& \kappa_4 \Lambda - \frac32 C \Lambda^2 \ ,\label{HBzeroM}\\
\H_1&=& \frac12 (\kappa_1 - 3 B \Lambda ) \sum_{k=1}^3 \left( x_k^2 + y_k^2 \right) \nn
&&
- \alpha x_1 - \beta_1 x_2 - \beta_2 (x_2 \cos \lambda + y_2 \sin \lambda) - \gamma (x_3 \cos \lambda + y_3 \sin \lambda),\label{HBunoM}\\
\H_2&=&  - \frac38 A \left(\sum_{k=1}^3 \left( x_k^2 + y_k^2 \right) \right)^2 .\ \label{HBdueM}
\eeqa
It is then useful to write the equations of motion 
\beqa{CEM}
\dot x_k &=& - \frac{\partial \H}{\partial y_k}, \;
\dot y_k =   \frac{\partial \H}{\partial x_k}, \label{CEM1} \\
\dot \Lambda &=& - \frac{\partial \H}{\partial \lambda}, \;
\dot \lambda = \frac{\partial \H}{\partial \Lambda}, \label{CEM2}
\eeqa
where the symplectic form
\beq{SF} \sum_{k=1}^3 d Q_k \wedge d q_k = \sum_{k=1}^3 d x_k \wedge d y_k,
\eeq
has been exploited, for each order term. Coherently with the book-keeping introduced above, we can write the series:
\beqa{solser}
\Lambda &=& \Lambda^{(0)} + \sigma^2 \Lambda^{(2)} + \dots, \\
\lambda &=& \lambda^{(0)}+\sigma^2\lambda^{(2)}+ \dots, \\ 
x_k &=& \sigma x_k^{(1)} + \sigma^3 x_k^{(3)} + \dots, \\
y_k &=& \sigma y_k^{(1)} + \sigma^3 y_k^{(3)} + \dots \ . 
\eeqa
At zero order, we get:
\beq{emP0}
\dot \Lambda^{(0)} =0, \quad \dot \lambda^{(0)} = \kappa_4 - 3 C \Lambda^{(0)}.
\eeq
At first-order, equations \eqref{CEM1} are:
\beqa{emP1}
\dot x_1^{(1)}&=& -(\kappa_1 - 3 B \Lambda^{(0)} ) y_1^{(1)} , \label{xdotP1} \\
\dot y_1^{(1)}&=& -\alpha + (\kappa_1 - 3 B \Lambda^{(0)} )  x_1^{(1)} , \\
\dot x_2^{(1)}&=& -(\kappa_1 - 3 B \Lambda^{(0)} )  y_2^{(1)} + \beta_2 \sin \lambda^{(0)} , \\
\dot y_2^{(1)}&=& -\beta_1  + (\kappa_1 - 3 B \Lambda^{(0)} )  x_2^{(1)} - \beta_2  \cos \lambda^{(0)} , \\
\dot x_3^{(1)}&=& -(\kappa_1 - 3 B \Lambda^{(0)} )  y_3^{(1)} + \gamma \sin \lambda^{(0)}, \\
\dot y_3^{(1)}&=& (\kappa_1 - 3 B \Lambda^{(0)} )  x_3^{(1)} - \gamma \cos \lambda^{(0)} , \label{qdotP1}
\eeqa
and at 2nd-order, we can consider only
\beqa{emP2}
\dot \Lambda^{(2)}&=& \beta_2 (y_2^{(1)} \cos \lambda^{(0)} - x_2^{(1)} \sin \lambda^{(0)}) + \gamma (y_3^{(1)} \cos \lambda^{(0)} - x_3^{(1)} \sin \lambda^{(0)}) \ , \label{emL1} \\
 \dot \lambda^{(2)}&=& - 3 C \Lambda^{(2)}  - \frac32 B \sum_{k=1}^3 \left( x_k^{(1)2} + y_k^{(1)2} \right) . \label{emL2} \eeqa

\subsection{De Sitter equilibria}\label{DSE}

Let us denote with $X_k,Y_k,\Lambda_E,\lambda_E$ the equilibrium values of the $x_k, y_k$ and the pair $\Lambda,\lambda$, assuming a series in $\sigma$ also for these quantities.

We immediately find a simple approximation of the de Sitter equilibria in the form of the zero-order solution to $\dot \lambda^{(0)} = 0$ provided by \eqref{emP0}
\beq{Q40}
\Lambda_E^{(0)} = \frac{\kappa_4}{3C}
\eeq
and to $ \dot x_k^{(1)} = \dot y_k^{(1)}=0$ given by \eqref{xdotP1}--\eqref{qdotP1}:
\beqa{eq0}
X_1^{(1)}&=& \frac{\alpha}{\omega} \ , \label{xzero1} \\ 
Y_1^{(1)} &=& 0 \ , \\
X_2^{(1)}&=& \frac{\beta_1  \mp \beta_2}{\omega} \ , \label{xzero2} \\
Y_2^{(1)} &=& 0 \ , \\
X_3^{(1)}&=& \mp\frac{\gamma}{\omega} \ , \label{xzero22} \\
Y_3^{(1)} &=& 0 \ , \label{xzero3}
\eeqa
with respectively $\lambda_E^{(0)}=\pi$ and $\lambda_E^{(0)}=0$ (we will see that the first case plays a special role). The new `slow' frequency 
\beq{F41}
\omega = \kappa_1 - 3 B \Lambda_E^{(0)} = \kappa_1 - \frac{B}{C} \kappa_4,
\eeq
has been introduced and readily, the first-order   correction for $\Lambda_E,\lambda_E$ is obtained from \eqref{emL1} and from \eqref{emL2}, 
with vanishing left-hand side at equilibrium: \rm
\beqa{eq1}
\Lambda_{E\mp}^{(2)}&=& - \frac{B \left(\alpha^2 + (\beta_1  \mp \beta_2)^2 + \gamma^2 \right)}{2C\omega^2}, \quad  \lambda_E^{(2)} = 0\ .
\eeqa
In these expressions, where different signs appear, they respectively correspond either to 
$\lambda=\pi$ (those with the upper sign) or to 
$\lambda=0$  (lower sign): we keep this  convention in all results obtained in the following. We have also to remark that 
$\omega$ is usually smaller than $\kappa_1, \kappa_4$: however, even smaller characteristic frequencies will appear in the process of normalization, so that we could better say that 
$\omega$ is \sl semi-slow. \rm 
  Moreover, we assume overall that $\omega$ does not vanish: this excludes the singular 
occurrence of exact resonance and is justified by the stability analysis implemented in the following subsection.  
The higher-order correction \rm $x_j^{(3)}$ can be obtained by solving the 3rd-order equations 
\beqa{emP3}
\dot x_j^{(3)}&=& - \omega y_j^{(3)} + \frac32 A y_j^{(1)} \sum_{k=1}^3 \left( x_k^{(1)2}+ y_k^{(1)2} \right),\\
\dot y_j^{(3)}&=& \omega x_j^{(3)} - \frac32 A x_j^{(1)} \sum_{k=1}^3 \left( x_k^{(1)2} + y_k^{(1)2} \right).
\eeqa
At equilibrium, they give
\beqa{eq2}
X_{k\mp}^{(3)}&=& \frac{3 (AC-B^2)}{2C\omega} X_{k\mp}^{(1)} \left(X_{1}^{(1)2} + X_{2\mp}^{(1)2} + X_{3\mp}^{(1)2}\right), \; k=1,2,3.\label{xuno1} 
\eeqa
These solutions can be interpreted as the equilibrium values of the $x_k, y_k$ providing the \sl forced eccentricities. \rm The zero-order terms (\ref{xzero1}--\ref{xzero3}) are dominant, so their sign allows us to identify the libration centers (if any) for the resonant combinations. Since the coupling parameters $\alpha,\beta_1,\beta_2,\gamma$ have definite signs \citep{BaMo2} for any reasonable combination of physical quantities, it is the sign of $\omega$ which produces different possibilities: this result agrees with what was already obtained by \citet{Si} and \citet{Si75} and in the following we will refer to these solutions as \sl de Sitter-Sinclair \rm equilibria. The updated solutions can therefore be written in the form 
\beq{Q41}
\lambda=\pi,0; \quad \Lambda_{E} = \frac{\kappa_4}{3C} -  \frac{B}{2C} \left(X_1^2 + X_2^2 + X_3^2 \right)
\eeq
where 
\be X_k = X_k^{(1)} + X_k^{(3)} , \, k=1,2,3. \label{X0N} \ee

We can however inquire about a possibility not included in this perturbative approach: one (or more) of the forced eccentricities can be so big to violate the book-keeping hierarchy assumed above. In this respect, we have to consider the case that also this quantity must be taken as a zero-order term in the book-keeping parameter   and we are required to 
take into account also 2nd-order terms in the eccentricities. These are described by the quadratic form \citep{HeCons,Ma,ShMa,CePaPuChao} \rm
$$
\H_{12} = - \sum_{\ell=-2}^2 \sum_{|n|+|m|=2} \alpha_{\ell nm} 
x^{n} y^m e^{i \ell \lambda}
$$
  where the multi-indexes $x^n = x_1^{n_1}x_2^{n_2}x_3^{n_3}, y^m = y_1^{m_1}y_2^{m_2}y_3^{m_3}$ are used 
and the $\alpha_{\ell nm}$ are coupling constants, first-order in the mass ratios, suitable generalisations of \eqref{PAR1}-\eqref{PAR4}. \rm 
In other words, we can look for additional solutions to the three equations 
\ba
\left(\omega -2 \alpha_2 + \frac{3(B^2-AC)}{2C} \left(X_1^2 + X_2^2 + X_3^2 \right) \right) X_1 - \alpha_{12} X_2 -\alpha &=&0, \label{NEQ1} \\
\left(\omega -2 \beta_{12} + \frac{3(B^2-AC)}{2C} \left(X_1^2 + X_2^2 + X_3^2 \right) \right) X_2 - \alpha_{12} X_1 \pm \gamma_2 X_3 -\beta_1 \pm \beta_2 &=&0, \label{NEQ2} \\
\left(\omega -2 \gamma_3+ \frac{3(B^2-AC)}{2C} \left(X_1^2 + X_2^2 + X_3^2 \right) \right) X_3 + \gamma_2 X_2 \pm \gamma &=&0, \label{NEQ3} \ea
where \eqref{Q41} has been inserted into the equilibrium conditions still considering the $X_k$ as unknown   and shorthand notation has been used for the non-vanishing coupling constants. \rm In order to test this possibility, let us assume that they exist additional equilibrium solutions with, e.g., $X_1 \gg X_2,X_3$. By using this condition in \eqref{NEQ1}, we get
\be
\left(\omega - K X_1^2 \right) X_1 - \sigma (\alpha + 2 \alpha_2 X_1) =0, \quad K \doteq \frac{3(AC-B^2)}{2C} \label{C3}
\ee
where now only   $\alpha$ and $\alpha_2$, defined as
\be\label{alfa2}
\alpha_2 \doteq \alpha_{0,200,000} = \frac{\overline m_2^2 \e_2 f_3 (\rho_{12})}{\overline L_1 \overline L_2^2} > 0, \quad 
\left(f_3(\rho) = {1\over 8}\left(44+14\rho{d\over{d\rho}}+\rho^2{d^2\over{d\rho^2}}\right)b_{1/2}^{(4)}(\rho)\right)
\ee
are assumed to be small parameters. \rm
This cubic can indeed have three real solutions if its discriminant is positive and they can   be explicitly written down \rm \citep{AL}. 
However, in view of the structure of the equation, we can proceed in a simpler way,  looking now for solutions of the form
$$
X_1=X_1^{(0)} + \sigma X_1^{(1)}.$$
To order zero in $\sigma$, we get the three solutions
$$
 X_1^{(0)} =0
$$
and
$$ X_1^{(0)} = \pm \sqrt{\frac{\omega-2\alpha_2 }{K}}.$$
The first of these provides again $X_1^{(1)} = \alpha/\omega$, namely \eqref{xzero1} already obtained above. 
But, if the argument of the square root is positive, we obtain two {\rm new solutions}:
\be
X_{1}^{(N)} = \pm \sqrt{\frac{\omega-2\alpha_2 }{K}} - \frac{\alpha}{2\omega}, \;\; N=2,3. \label{X1N} \ee
Since,   recalling (\ref{Adef}--\ref{Cdef}) and the definition in (\ref{C3}), $K$ is always positive, \rm the necessary condition for these new solutions is strictly
\be
\omega = \kappa_1 - \frac{B}{C} \kappa_4 > 2 \alpha_2 > 0, \label{condX1N}\ee
whereas we recall that the de Sitter-Sinclair solution allowed for both signs of $\omega$. 
With this result plugged in the other two equations \eqref{NEQ2} and \eqref{NEQ3} (still under the assumption $X_1 \gg X_2,X_3$) we get the new solutions 
\be
X_{2}^{(N)} = \frac{\beta_1  \mp \beta_2 + \alpha_{12} X_{1}^{(N)}}{\omega -2 \beta_{12} - K (X_{1}^{(N)})^2}  \label{X2N} \ee
and
\be
X_{3}^{(N)} = \frac{\mp \gamma}{\omega -2 \gamma_3 - K (X_{1}^{(N)})^2} \label{X3N} \ee
to be respectively added to  \eqref{xzero2} and  \eqref{xzero22}. In every cases, the equilibrium value of the $Y_k$ is still zero. 
  The relevant coupling constants appearing in the new solutions are \rm
\ba\label{alfak}
\alpha_{12} &\doteq& \alpha_{0,110,000} = \frac{\overline m_2^2 \e_2}{\sqrt{\overline L_1} \overline L_2^{3/2}} (f_5+f_6) (\rho_{12}), \\
\beta_{12} &\doteq& \alpha_{0,020,000} + \alpha_{2,020,000}=
\frac{ \overline m_2^2 \e_2 f_4 (\rho_{12})}{\overline L_2^{3}} + \frac{\overline m_2 \overline m_3^2 \epsilon_3 f_3 (\rho_{23})}{\overline L_3^2 \overline L_2}, \\
\gamma_{3} &\doteq& \alpha_{0,002,000} =\frac{\overline m_2 \overline m_3^2 \epsilon_3 f_4 (\rho_{23})}{\overline L_3^3}, \\
\ea
  where $f_3$ is defined in \eqref{alfa2} and \rm
\ba\label{betak}
f_4(\rho) &=&   {1\over 8}\left(38+14\rho{d\over{d\rho}}+\rho^2{d^2\over{d\rho^2}}\right)b_{1/2}^{(2)}(\rho), \\
f_5(\rho) &=&  -{1\over 4}\left(42+14\rho{d\over{d\rho}}+\rho^2{d^2\over{d\rho^2}}\right)b_{1/2}^{(3)}(\rho),\\
f_6(\rho) &=&   {1\over 4}\left(2-2\rho{d\over{d\rho}}-\rho^2{d^2\over{d\rho^2}}\right)b_{1/2}^{(1)}(\rho).
\ea
We remark that exact solutions of the cubic can be explicitly computed. However, this would still be an incomplete representation of the whole set whose 
algebraic expression is very involved. Moreover, in order to not overload the notation, we have considered together solutions corresponding to both 
$\lambda=\pi$ and $\lambda=0$. In practice, a direct numerical solution of the equilibrium equations will be performed specifying to which kind of equilibrium one is 
referring to. What is important now is to highlight the existence of additional possible equilibria with respect to the well known de Sitter-Sinclair ones. 
We will refer to these new equilibria as 
\sl new de Sitter \rm solutions. We will see later if they actually play some relevant role.

\subsection{Stability of the de Sitter equilibria} 
\label{sec:stbequi}

A stability analysis of these equilibria can be performed with standard techniques and compared with 
other numerical \citep{HM81,YoPe} and analytical \citep{Si75,ZHB} studies. Collectively denoting with $z$ the set $(x_k, y_k,\Lambda,\lambda)$, 
we consider the linear Hamiltonian equations providing the variational system \citep{YoPe}
\be\label{VES}
\dot \delta z = {  J } H_{zz} \big\vert_0 \delta z.\ee
Looking for solutions of the form
$$
\delta z = Z e^{\mu t} $$
we have to compute the eigenvalues of the Hamiltonian matrix in \eqref{VES}.  
The case of the first class of de Sitter-Sinclair equilibria \eqref{Q40}-\eqref{xzero3} can be treated explicitly. 
We have to compute the eigenvalues of the matrix 
\be\label{MVES}
{  J } H_{zz} \big\vert_0 = 
\left(\begin{array}{cccccccc}
	0 & 0 & 0 & 0 & -\omega & 0 & 0 & 0 \\
	0 & 0 & 0 & 0 & 0 & -\omega & 0 & \mp \beta_2  \\ 
	0 & 0 & 0 & 0 & 0 & 0 & -\omega & \mp \gamma \\ 
	0 & 0 & 0 & 0 & 0 & \mp \beta_2 & \mp \gamma & \frac{\pm \beta_1 \beta_2 - \beta_2^2 - \gamma^2}{\omega} \\ 
	\omega & 0 & 0 & -\frac{3B \alpha}{\omega} & 0 & 0 & 0 & 0 \\ 
	0 & \omega & 0 & -\frac{3B (\beta_1 \mp  \beta_2)}{\omega} & 0 & 0 & 0 & 0 \\ 
	0 & 0 & \omega & \pm \frac{3B \gamma}{\omega} & 0 & 0 & 0 & 0 \\
	-\frac{3B \alpha}{\omega} & -\frac{3B (\beta_1 \mp  \beta_2)}{\omega} & \pm \frac{3B \gamma}{\omega} & -3C & 0 & 0 & 0 & 0 \\
                 \end{array}\right).
\ee
Like before, where different signs appear, they respectively correspond either to $\lambda=\pi$ (upper sign) or to $\lambda=0$  (lower sign). The eigenvalues $\mu_a, \, a=1,\dots,8,$ are the 
solutions of the characteristic equation
$$
\det ( {  J } H_{zz} \big\vert_0 - \mu {  I }) = 0$$
which has the form
\beqa{CEQ}
&& \frac1{\omega^2} (\mu^2+\omega^2) [ -6 B (\beta_1 \beta_2 - \beta_2^2 - \gamma^2) \mu^2 \
\omega(\mu^2 + \omega^2) + \nonumber \\ 
&& \omega (\mu^2 + \ \omega^2) (-3 C (\beta_2^2 + \gamma^2) \mu^2 + 
    3 C \beta_1 \beta_2 (\mu^2 + \omega^2) + \mu^2 \ \omega (\mu^2 + \omega^2)) + \nonumber \\
&& 9 B^2 (\alpha^2 (-(\beta_2^2 + \gamma^2) \mu^2 + \beta_1 \ \beta_2 (\mu^2 + \omega^2)) + \beta_1 (\beta_1^2 \beta_2 \
(\mu^2 + \omega^2) + \nonumber \\ 
&& \beta_2 (\beta_2^2 + \gamma^2) (\
\mu^2 + \omega^2) - \beta_1 (\gamma^2 \mu^2 + 
          2 \beta_2^2 (\mu^2 + \omega^2))))] = 0.
\eeqa
The explicit expression of the solution is too cumbersome to be reproduced here. However, the stability property can still be described working only with the determinant 
of the matrix itself which is
\be\label{MES}
\pm 3 \beta_1 \beta_2 \omega^2 (3 B^2 (\alpha^2 + (\beta_1 \mp \beta_2)^2 + \gamma^2) + C \omega^3).
      \ee
      A pair of eigenvalues $\pm i \omega$ is already factored out in \eqref{CEQ}.   Three other pairs remain, \rm which, for a Hamiltonian matrix, are either pure imaginary or 
real and opposite. Therefore a sufficient condition for instability is that the determinants are negative: in practice, this condition is also necessary, because only one pair 
of real eigenvalues appear in standard cases. By using \eqref{xzero1}-\eqref{xzero3} the determinants in the two cases can be rewritten as 
$$
\pm 3 \beta_1 \beta_2 \omega^4 (3 B^2 (X_1^2 + X_2^2 + X_3^2) + C \omega)
      $$
      It is straightforward to check that $ \beta_1 \beta_2 <0$ for every reasonable choice of parameters, therefore a sufficient condition for instability is
      \beq{LUNST}
      \omega > \omega_U \doteq - \frac{3B^2}C (X_1^2 + X_2^2 + X_3^2)
     \eeq
     in the first case ($\lambda=\pi$) and $\omega < \omega_U$ in the second case. We remark that these findings agree with \citet{Si75} results. 
     The analytical evaluation of the instability transition in the case of the new-de Sitter equilibria 
     is much more involved. However, direct numerical computation of the eigenvalues can be easily performed to predict the stability properties of the additional solutions.

\section{Normal forms} 
\label{sec:nfs}

\subsection{Previous attempts at normalisation} 
\label{sec:modelsecondary}

A 4-DOF multi-resonant system shows intricate dynamics. 
There have been some previous works producing simplified models. 
We first mention the works by \citet{FeBook}, \citet{Ma} and \citet{ShMa}, 
in which a quite general model for the Galilean system (including dissipation) is presented and solved in terms of `variational' solutions. 
Averaging methods have been henceforth applied in several variants \citep{BDH,MCB,L18}. However, they always result in non-integrable systems 
and, in the end, worth of numerical evaluations only. 

A more effective approach is that of constructing a `normal form' which captures 
the resonant dynamics by means of a near-identity canonical transformation. This is usually generated 
by enforcing the commutation of the new Hamiltonian with a predefinite integrable part. 
With suitable assumptions and restrictions, integrability can be extended to the normal form itself. 
Extending integrability breaks when more than one exact resonances are present,   
although an approximate integrable Hamiltonian can be constructed in some cases \citep{Hadden19}, so
even in the multi-resonant case \rm a resonant normal form provides several useful informations.

A remarkable attempt is due to \citet{He}  who constructed a normal form by eliminating all angles but the Laplace argument $\lambda$. His aim was to have an analytical tool to explore models of capture in which the libration of the Laplace argument may not necessarily be small. We observe that, even without an explicit statement, Henrard's normal form in \citep{He} is constructed by expanding around the de Sitter equilibrium. Analogously, in \citep{CePaPu,CePaPuAA} we have constructed a normal form in which all angles are eliminated, except $q_3$: this choice was dictated by the objective of enlightening the apparently important role played by this angle in determining the nature of the dynamics. However, the resonant dynamics are effectively better described when considering the $\Lambda-\lambda$ plane and therefore here we pursue Henrard's approach but with two important differences: the use of the modified variable set and a completely symbolic approach not limited to the numerical data of the Galilean system.

\subsection{Laplace normal form}
Once obtained the two sets of equilibria with 
$$
\lambda=0,\pi; \quad \Lambda=\Lambda_{E\mp},$$
we can address the investigation of the dynamics around them. The best strategy for this is to normalise the Hamiltonian by eliminating `fast' angles. In analogy with Henrard's approach, 
we do not limit the construction to the neighbourhood of the equilibrium but allow for large amplitude librations. Therefore, considering as a reference the `standard' approximate $\pi,\Lambda_E^{(0)}$ equilibrium provided by \eqref{Q40}, we only perform the translation 
$$\Lambda \rightarrow \hat\Lambda = \Lambda - \Lambda_E^{(0)} = \Lambda - \frac{\kappa_4}{3C}.$$ 
The model Hamiltonian can then be rewritten as 
\beq{resoH}
H = (\omega - 3 B \hat\Lambda )  \Gamma - \frac32 C \hat\Lambda^2 - \frac32  A \Gamma^2 + H_{res} (Q_1,Q_2,Q_3,q_1,q_2,q_3^M,\lambda) ,\eeq
where $\Gamma$ is the angular momentum deficit introduced in \eqref{amddef} and 
$H_{res}$ is the resonant coupling part of \eqref{HMdueM}.

The $\omega$ frequency is associated with the free eccentricity oscillations. We assume it to be `fast' with respect to the libration of the Laplace argument. We can therefore normalise with respect to the `isotropic oscillator'
$$
H_I=\omega \Gamma = \omega (Q_1+Q_2 + Q_3)
$$
by removing the dependence of the Hamiltonian on fast angles. Resonant combinations cannot be removed and actually produce interesting phenomena like `beats' in the eccentricities. However, they appear only at order higher than two and, in case, they can be included by imposing equilibrium values for resonant angles different from $\lambda$. A preliminary analysis of their role is performed in the following subsection.

For sake of completeness we briefly recall some aspects of resonant normalisation. A given set of frequencies $\kappa_a, a=1,\dots,n$ is \sl resonant \rm if
\beq{resor}
\sum_{a=1}^n \ell^{\ (b)}_a \kappa_a = 0, \; \ell_a \in \Z^n , \; b=1,\dots,m \ . \eeq
The vectors $\{{\vec \ell}^{\ (b)}\}$ provide the \sl resonant module; \rm the minimum of $||{\vec \ell}^{\ (b)}||-1, \, b=1,\dots,m$ gives the \sl order \rm of the resonance. $H_I$ is fully isotropic in the space of frequencies, so $m=3$ and the first-order resonant module is given by the vectors:
\beq{rmviso}
{\vec \ell}^{\ (1)} = (1,-1,0), \quad
{\vec \ell}^{\ (2)} = (1,0,-1), \quad
{\vec \ell}^{\ (3)} = (0,1,-1).
\eeq

We construct the resonant normal form by enforcing the commutation of the terms in the new Hamiltonian with $H_I$. 
The `Laplace normal form' is constructed by implementing a Lie transform with resonant module spanned by the vectors \eqref{rmviso} \citep{cefthym}. Using primes to denote the new variables, the first-order normalization gives the following Hamiltonian
\beq{nfl}
K = \omega  (Q'_1+Q'_2 + Q'_3) - 3 B \hat\Lambda' (Q'_1+Q'_2 + Q'_3) - \frac32 C \hat\Lambda'^2 - \frac32  A (Q'_1+Q'_2 + Q'_3)^2 - \frac{\beta_1 \beta_2}{\omega} \cos \lambda'.
\eeq
  The construction around the other equilibrium $\lambda=0$, would lead to the same function 
with the positive sign in front of the cosine term. \rm
The transformed angular momentum deficit can be denoted as
\beq{amddefp}
\Gamma'=Q'_1+Q'_2 + Q'_3,
\eeq
and, in this approximation, is a conserved quantity. We are therefore led to investigate the reduced Laplace Hamiltonian
\beq{rnfl}
K_L = 3 B \Gamma' \hat\Lambda'  + \frac32 C \hat\Lambda'^2 + \frac{\beta_1 \beta_2}{\omega} \cos \lambda'.
\eeq
It provides a first-order approximation of the libration frequency around the reference $\lambda=\pi$ equilibrium
\beq{fola}
\omega_L=\sqrt{\frac{3C\beta_1 \beta_2}{\omega}}\eeq
and an approximate resonance width given by
\beq{lrw}
\Delta \Lambda=4 \sqrt{\frac{\beta_1 \beta_2}{3 C \omega}}.\eeq
  These results are analogous to those derived by \citet{QA} in her general treatment of pure three-body resonances. \rm

The dynamics of the non-linear oscillators is given by
$$
\dot Q'_k = 0, \quad \dot q'_k = \frac{\partial K}{\partial Q'_k} = \omega_\mathrm{free},$$
where the frequency 
\beq{FF}
\omega_\mathrm{free} = \kappa_1 - 3 \left(B \Lambda + A \Gamma' \right)
\eeq
gives the free eccentricity oscillations. $\omega$ is therefore the first order approximation of 
$\omega_\mathrm{free}$ at the equilibrium value $\Lambda=\Lambda_E$. 
It is worthwhile to remark that, the smaller the value of $\omega$ the larger the resonance width. 
On the other hand, a lower bound for $|\omega|$ is provided 
by \eqref{LUNST}. We can add a general result by observing that, 
when inserting in \eqref{FF} the appropriate definitions and the equilibrium results obtained above, 
we get
\beq{FF2}
\omega_\mathrm{free} = \frac3C \left[(3B -2C) \eta_2 L_2 - (B -C) \eta_1 L_1 - 2B \eta_3 L_3\right].
\eeq
We see that, by choosing exactly resonant nominal values, namely such that $\overline n_1 = 2 \overline n_2 = 4 \overline n_3$,
$\omega_\mathrm{free}$ vanishes and so does $\omega$ if the nominal choice $Q'_k=0$ is done. 

Higher-order evaluations of these quantities can be performed by means of higher-order normal forms. 
Explicit algebraic expressions are cumbersome and will be reported in a forthcoming publication \citep{ckpv}. 
Numerical values for a more accurate modelling of the Laplace libration in the Galilean system are provided in the Appendix. 
It is interesting to compare this normal form with that obtained in the pioneering work by \citet{He}. 
Apparently, Henrard normal form was constructed by suppressing free eccentricities 
(see the next subsection) and therefore locating the system in a de Sitter equilibrium. 
This clearly does not prevent an accurate evaluation of the libration frequency of the 
Laplace argument but leads to an incomplete description of all the aspects of the dynamics of the Laplace resonance.

\subsection{Forced and free eccentricity oscillations}
\label{sec:freefo}
Let us now collectively denote the canonical coordinates as $W=(Q,\Lambda,q,\lambda)$. The original coordinates are given by a series of the form $\sum_{k} \sigma^k W_k$ such that, 
in terms of the new normalizing coordinates, they are given by
\beqa{solss}
W_0 &=&  W', \\
W_1 &=& \{W',\chi_1\}, \\
W_2 &=& \{W',\chi_2\} + \frac12 \{\{W',\chi_1\},\chi_1\}, \\
\vdots &=& \vdots
\eeqa
The first two generating functions of the Laplace normal form are 
$$
\chi_1 = -\frac1{\omega}
\left[\alpha \sqrt{2 Q'_1} \sin{q'_1} +
\beta_1 \sqrt{2 Q'_2} \sin{q'_2} + \beta_2 \sqrt{2 Q'_2} \sin{(q'_2-\lambda')} + \gamma \sqrt{2 Q'_3} \sin{(q'_3-\lambda')} \right]
$$
and $\chi_2=0$, so that, to second order we get
\beqa{emP1t}
x_1&=& x'_1 - \frac{\partial \chi_1}{\partial y'_1} = x'_1 + \frac{\alpha}{\omega}, \label{xFFP1} \\
y_1&=& y'_1 + \frac{\partial \chi_1}{\partial x'_1} =  y'_1, \\
x_2&=& x'_2 - \frac{\partial \chi_1}{\partial y'_2} = x'_2 + \frac1{\omega} \left( \beta_1 + \beta_2 \cos \lambda' \right), \\
y_2&=& y'_2 + \frac{\partial \chi_1}{\partial x'_2} =  y'_2 +  \frac{\beta_2}{\omega} \sin \lambda' , \\
x_3&=& x'_3 - \frac{\partial \chi_1}{\partial y'_3} = x'_3 + \frac{\gamma}{\omega} \cos \lambda', \\
y_3&=& y'_3 + \frac{\partial \chi_1}{\partial x'_3} =  y'_3 + \frac{\gamma}{\omega} \sin \lambda' , \label{xFFP3}
\eeqa
where the free eccentricity oscillations are 
$$
x'_k = a_k \cos \omega_\mathrm{free} (t-t_0), \quad y'_k = a_k \sin \omega_\mathrm{free} (t-t_0)$$
and the amplitudes $a_k$ are fixed by initial conditions. 

These relations show how the transformation to the Laplace normal form, aimed at removing 
non-resonant terms 
depending on $q_k$, automatically shifts the eccentricity vectors to the forced equilibria (apart 
for the slow modulation due to $\lambda$), a nice feature already exploited in the approach by \cite{He}.

\subsection{Higher-order Resonant normalisation}
\label{sec:reso}
The resonant normal forms allows us to investigate additional slow phenomena associated with beats among resonant terms. To evaluate them, the model \eqref{HMzero}--\eqref{HMdue} expressed in the original Henrard-Malhotra coordinate performs better because the frequency vector is not degenerate, breaking the isotropy of the linear oscillator. 
For the frequencies appearing in $H_0$ (see \eqref{HMzero}), the resonant module is now given by the two vectors:
\beqa{rmv}
{\vec \ell}^{\ (1)} &=& (k,-k,0,0), \quad k \in \N \ ,\\
{\vec \ell}^{\ (2)} &=& (k_1,k_2,k_3,k_3), \quad k_1+k_2=-k_3, \quad {  k} \in \N^3 \ .
\eeqa

The procedure based on the Lie transform approach now produces the following Hamiltonian (again, for sake of simplicity, we leave the same symbols for the new normalising variables):
\beq{KLP}
K (Q_a,q_a)=\sum_{k=0}^2 K_k  (Q_a,q_a), \quad  a=1,...,4 \ ,
\eeq
with
\beqa{KFS}
K_0&=& H_0 + c_1 Q_1 + c_2 Q_2 + c_3 Q_3 + c_4 \widehat Q_4 \ ,\label{KMzero}\\
K_1&=& H_1 \ ,\label{KMuno}\\
K_2&=& C_{12} \sqrt{Q_1 Q_2} \cos (q_1 - q_2) + C_{23} \sqrt{Q_2 Q_3} \cos (q_2 - q_3 - q_4)\ .\label{KMdue}
\eeqa
In the linear part, the frequencies appear to be amended by the following coefficients:
\beqa{cn}
c_1 &=& -\frac{3 (2 \alpha^2 + \beta_1^2) (\eta_1 + 4 \eta_2)}{2 \kappa_1^2} + \frac{3(\beta_2^2 + \gamma^2) \eta_2}{\kappa_3^2} \ , \\
c_2 &=& \frac{3 \alpha^2 \eta_2}{\kappa_1^2}-\frac{3\beta_1^2(\eta_1 + 2 \eta_2)}{2 \kappa_1^2}-\frac{3(2\beta_2^2 + \gamma^2) (\eta_2 + 4 \eta_3)}{2 \kappa_3^2} \ , \\
c_3 &=& \frac{3(\alpha^2 + \beta_1^2) \eta_2}{\kappa_1^2}-\frac{3 (\beta_2^2 + 2\gamma^2) (\eta_2 + 4 \eta_3)}{2 \kappa_3^2} \ , \\
c_4 &=& -\frac{3(\alpha^2 + \beta_1^2) (\eta_1 + 6 \eta_2)}{2\kappa_1^2}+\frac{3 (\beta_2^2 + \gamma^2) (3\eta_2 + 4 \eta_3)}{2 \kappa_3^2} \ . 
\eeqa
The quadratic part has the same form \eqref{HMuno} as in the original Hamiltonian. The resonant part is characterised by the two resonant combinations $q_1 - q_2$ and $q_2 - q_3 - q_4$ and the two coefficients
\beqa{cnn}
C_{12}  &=& -\frac{3 \alpha \beta_1 (\eta_1 + 4 \eta_2)}{\kappa_1^2} \ , \\
C_{23}  &=& -\frac{3 \beta_2 \gamma (\eta_2 + 4 \eta_3)}{\kappa_3^2} \ ,
\eeqa
The simplification offered by the normal form is evident when considering that the dimensionality of the system is reduced from 4 to only 2 degrees of freedom. In fact, we recognise the existence of the two additional formal integrals
\beq{nfi}
\E = Q_3 - \widehat Q_4 
\eeq
and $\Gamma$, the angular momentum deficit already introduced in \eqref{amddef}.

By means of the canonical transformation $(Q_a, q_a) \longrightarrow (R_1,R_2,\E,\Gamma,r_1,r_2,e_1,e_2)$
\ba\label{cootrNF}
r_1&=&q_1-q_2 \ , \quad\quad\quad\quad \, R_1 = Q_1 \ ,\label{qn1}\\
r_2&=&q_2-q_3-q_4 \ , \quad\quad R_2 = Q_1+Q_2 \ , \label{qn2}\\
e_1&=&-q_4 \ , \quad\quad\quad\quad\quad \;\; \E = Q_3 - \widehat Q_4 \ , \label{qn3}\\
e_2&=&q_3+q_4 \ , \quad\quad\quad\quad \,
\Gamma = Q_1 + Q_2 + Q_3 \ , \label{qn4}\ea
 the resonant normal form can be written as 
\beqa{nfnris}
K_R (R_1,R_2,r_1,r_2;\Gamma)&=& C_1 R_1 + C_2 R_2 + \nonumber \\
&&C_{12} \sqrt{R_1 (R_2-R_1)} \cos r_1 + \nonumber \\ &&C_{23} \sqrt{(\Gamma -R_2) (R_2-R_1)} \cos r_2 \ , 
\eeqa
where
\beqa{nfreq}
C_1 &=& c_1 - c_2 \ ,\label{omeuno}\\
C_2 &=& c_2 -c_3 -c_4\ .\label{omedue}
\eeqa

\n
The normal form $K_R (R_1,R_2,r_1,r_2)$ can be used to investigate the dynamics on longer time-scales than those associated with the libration of the Laplace arguments. For example, in the case of the Galilean satellites, on time-scales of the order of $50000 \div 100000$ days, on which we see modulations of the eccentricities due to the resonance. On the other side, we deduce that the Laplace normal form model of the previous subsection, implementing the inverse transformation in Poincar\'e variables, is good for the short-time dynamics of the libration of the Laplace arguments.

\subsection{Detuning the exact resonant dynamics}\label{DNF}
Secular precessions (e.g. due to the quadrupole) break the exact resonance. 
However, non-linear coupling restore the resonant behaviour slightly changing libration frequencies. 
In order to describe additional precessional effects, we can add to $H_{Kep}$ a secular part depending on the $Q_k$ 
to include in the model the effects due to the oblateness of the central body and, possibly, the averaged motion of other bodies (`Callisto'). 
The main effects due to the quadrupole of the central body can be described by a term
\be\label{Hsec_esp}
H_{sec}= - \frac32 J_2 \sum_{k=1}^3 (R_J / a_k)^2 n_k Q_k \ ,
\ee
where $J_2$ is the quadrupole coefficient and $R_J$ is the radius of the central body (`Jupiter').

In place of \eqref{HMzeroM}, the linear part of the expansion can now be written as
\be
H_0^{D} = \sum_{j=1}^3 \kappa_j Q_j + \kappa_4^M \Lambda ,\label{HMzerodet}
\ee
where
\be
\kappa_j = \kappa_1^M + \frac{\partial H_{sec}}{\partial Q_j}, \label{kappadet}
\ee
are the detuned frequencies. The corrections are usually small, therefore the normalization can be implemented 
in the same way as above by keeping resonant combinations in a detuned resonant normal form \citep{PBB}. 
The unperturbed part is a slightly anisotropic oscillator of the form
\beq{DF41} \sum_{j=1}^3 \omega_j Q_j  = \sum_{j=1}^3 \left(\kappa_j - \frac{B}{C} \kappa_4 + \frac{3(B^2-AC)}{C} \Gamma \right) Q_j , 
\eeq
where the semi-slow frequencies are defined as natural generalisations of \eqref{FF2}. As a 
meaningful example, in the Appendix is discussed the Galilean case when the oblateness of Jupiter is 
included in the model.


\section{Applications}
\label{SAPP}

In the present section we substantiate the results obtained above. First we give a resume of the outcomes of the simplified model and the associated normal form. Afterwards, we apply those results to the two most representative systems: the Galilean satellites of Jupiter and the extra-solar system GJ-876. Happily, these two examples are in a certain sense paradigmatic since each of them represents a general type of Laplace configurations.

\subsection{Summary of the results}
We can summarise the main outcome of the previous sections by recalling the main ingredients of the Laplace resonance dynamics. We assume an ideal system composed of a main spherical body $m_0$ and three point masses $m_j$ located on three nominal orbits assuring mean motion resonance, fixing in this way the semi-major axes. An analogous approach is usually adopted in investigating two-body mean-motion resonances \citep{BaMo,BaMo2}. Recalling \eqref{amg} and \eqref{ams}, the system parameters are therefore:
$$
\epsilon_j, \quad j=1,2,3; \quad \overline m_A, \quad \bar a_A = \frac{a_A}{a_1}\ , \quad A=2,3,$$
which give:
\beqa{LN}
\overline L_1&=& \frac1{\sqrt{1+\epsilon_1}} \ ,\label{LN1}\\
\overline L_2&=& \frac{\overline m_2 (1+\epsilon_1)}{\sqrt{1+\epsilon_1+\epsilon_2}} \sqrt{\bar a_2}\ ,\label{LN2}\\
\overline L_3&=& \frac{\overline m_3 (1+\epsilon_1+\epsilon_2)}{\sqrt{1+\epsilon_1+\epsilon_2+\epsilon_3}} \sqrt{\bar a_3}\ .\label{LN3}
\eeqa
In this way, in view of \eqref{neta}, mean motions $n_j$ and $\eta_j$ are specified determining in turn the $A,B,C$ introduced in \eqref{Adef}-\eqref{Cdef} and the coupling constants defined in \eqref{PAR1}-\eqref{PAR4}. We remark that these nominal values are used as `seeds' to compute reference quantities specifying the models. In particular, the trivial choice $\overline P_j = 0, j=1,2,3$   produces in practice reliable models as a more accurate choice  would do when based on actual elements. \rm However, this choice combined with the nominal resonant $\overline L_j$ given above, would give, recalling \eqref{FF2}, a vanishing value for the frequency of the free eccentricity and a singularity in the Laplace libration. Since the model is completely defined by specifying the values of the two conserved momenta $Q_5, Q_6$ of \eqref{q5}-\eqref{q6}, we can chose initial conditions in a domain which is implicitly defined 
by $|\omega|>\omega_U$ to comply with \eqref{LUNST} but small enough to get a finite size for the resonance width. 

Osculating values of $L_j, Q_j$ are then obtained by exploiting the results of sections \ref{sec:bke} and \ref{sec:nfs}. The forced eccentricities are computed by using the solutions \eqref{X0N} or/and \eqref{X1N}-\eqref{X3N} to get
\beq{FECC}
Q_j^{FE (N)} = \frac12 (X_j^{(N)})^2, \quad N=1,2,3,
\eeq
and the libration centers can be evaluated by looking at the sign of the $X_j$ so that:
\beqa{LC}
q_1^{FE} &=& \arccos {sign{[X_1]}},\\
q_2^{FE} &=& \arccos {sign{[X_2]}},\\
q_3^{FE} &=& \arccos {sign{[X_3]}} - q_4.
\eeqa
The equilibrium  solution \eqref{Q41} and the libration frequency \eqref{fola} of the Laplace argument can henceforth be obtained. Finally, by using the forced values in \eqref{FECC}, the libration width \eqref{lrw} can be evaluated. 



\subsection{The case of the Galilean system}\label{GTM}
As a benchmark for these results we use an idealised version of the Galilean system, 
namely a mock-up of the actual system of the satellites of Jupiter 
with which to compare our predictions. We remark that this exercise has a pure 
pedagogic value and is not aimed at an accurate reconstruction of the system. 
It is well known that, for a good description of the Galilean satellite dynamics, 
we have to include at least the oblateness of the planet and to expand the disturbing function 
up to second order in the eccentricities \citep{YoPe,He,Ma,CePaPuAA}. However, for sake of understanding the qualitative aspects, 
the present simplified model is sufficient: for more accurate quantitative 
results we refer to the detuned resonant normal form illustrated in the Appendix.

In Table \ref{T1} we can see a comparison between nominal actual element values of the 
Galilean system and corresponding values obtained with the procedure outlined above: 
as said, this is just to have an idea of how far our toy model is from the actual system. 
Using these values, the two frequencies turn out to be
\beq{Gfreq}
\omega_\mathrm{free}=-0.0033, \quad \omega_L=0.0011,\eeq
taking into account the time scale implicit in the choice of unit 
($\omega=1/T$, with time unit = 1.7714 days).  They respectively correspond to the periods
\beq{GP1}
T_\mathrm{free}=536 \;\, days; \quad 
T_L=1584 \;\, days.\eeq
These values are in very good agreement with those obtained from numerical solutions with the toy model 
(for more accurate values concerning the `true' Galilean system, see Table \ref{T4}). 
The equilibrium value at the center of libration of $\Lambda$ computed with \eqref{Q41} is
\beq{GLL}
\Lambda=0.010087.
\eeq
The value corresponding to nominal values is $ 0.0100872 $. The resonance width (in agreement with \eqref{lrw}) is 
$$
\Delta \Lambda= 0.00022.$$
This result shows how small is the domain of the resonance and provides a strong condition for the architecture of the system. 
In fact, we can conjecture that, for 
some physical reason (e.g. dissipative long-term effects due to tidal interactions), 
the semi-major axes could be quite different from those observed but the combination 
among the $L_j$-s given by $\Lambda$ remains close to the equilibrium value. 
For sake of completeness, we provide a preliminary evaluation of the forced eccentricity 
when considering also the contribution of the quadrupole of Jupiter and expanding up to second order in the eccentricities \citep{ckpv}. 

\begin{table}

    \begin{tabular}{@{}llllcccc@{}}
  \hline\hline
semi-major axis [in unit of $a_1$]                                   	&\rm{Io}            & \rm{Europa}    & \rm{Ganymede} & & & & \\
\hline

 nominal							                    	&  1                  	& 1.5905            &       2.5365    & & & & \\
 de Sitter-Sinclair (analytical)                             			&  1                   & 1.5908            &   	 2.5366   & & & & \\
 de Sitter-Sinclair (analytical with $J_2$)				&  1                  	& 1.5905            &       2.5365    & & & & \\
 \hline\hline
 eccentricity/resonant angles 						& $e_1$		& $e_2$		& $e_3$ 	& $q_1$ \quad & \quad $q_2$& $q_3$ & $q_4$ \\
 \hline
 
nominal	                         										&  0.0042          & 0.0094           &       0.0015   & 0 \quad & \quad $\pi$ & rotating & $\pi$ \\
de Sitter-Sinclair (analytical)                                  					&  0.0058          & 0.0118           &       0.0008   & 0 \quad & \quad $\pi$ & $\pi$ & $\pi$ \\
de Sitter-Sinclair (analytical with $J_2$)                                			&  0.0042          & 0.0095	        &       0.0011   & 0 \quad & \quad $\pi$ & $\pi$ & $\pi$ \\
anti-de Sitter-Sinclair (analytical)                                					&  0.0055          & 0.0113	        &       0.0008   & 0 \quad & \quad 0 & $\pi$ & 0 \\
\hline\hline
\end{tabular}
\caption{Mean nominal orbital elements of the Galilean satellites according to 
\citet{La-A} compared with the predictions from the toy-model (and the upgraded model described in the Appendix) and libration centers of the resonant angles.} \label{T1}
\end{table}

Using the first-order solution \eqref{X0N}, the standard de Sitter equilibrium 
\beq{GDS}
q_1 = 0,\quad
q_2 =\pi , \quad
q_3 =\pi , \quad q_4 =\pi
\eeq
is stable: using \eqref{LUNST} we get $\omega_U = -0.00064$, which can be compared with the numerical outcome by \cite{YoPe} that, 
when converted to our units, is $-0.00068$. The eigenvalues of the matrix \eqref{MVES} of the variational system are in fact 
$$
\pm i \ 0.0011, \quad
\pm i \ 0.0024, \quad
\pm i \ 0.0032, \quad
\pm i \ 0.0033. 
$$
Their product gives the determinant \eqref{MES} in the case with upper signs.
They give two frequencies very close to those above in \eqref{Gfreq} and two others with periods of $\sim 540$ and $\sim 735$ days which can be traced in the evolution of the eccentricities of the toy model. We observe that, in this framework, the rotation regime of $q_3$ is possible if the amplitude of the free eccentricity of Ganymede is sufficiently high. 
It is worthwhile to remark that, even in this case (which is what happens in the real system), the libration regime is still possible \citep{LSF}. Therefore the libration around the de Sitter-Sinclair equilibrium can be a feasible outcome of the future long-term evolution of the Galilean system.

The ``anti-de Sitter'' equilibrium 
\beq{GADS}
q_1 = 0,\quad
q_2 = 0 , \quad
q_3 =\pi , \quad q_4 = 0
\eeq
is unstable: the eigenvalues of the matrix of the variational system (their product gives the determinant \eqref{MES} with lower signs) are now
$$
\pm 0.0021, \quad
\pm i \ 0.0039, \quad
\pm i \ 0.0040, \quad
\pm i \ 0.0047. 
$$
As a matter of fact, condition \eqref{condX1N} is never satisfied so the additional solutions \eqref{X1N}-\eqref{X3N} do not apply in the context of Galilean dynamics.

\subsection{The case of the GJ-876 system}\label{JTM}
The system Gliese-Jahrei{\ss} 876 (GJ-876 hereafter) plays a very important role in the now 25-years long history of exoplanets studies. It was the first case of detection of mean motion-resonance outside our Solar System 
\citep{marcyeal} and the first instance of multi mean motion-resonance among three planets \citep{RG876}. It is very close (4.689 parsec) and therefore radial velocity and photometric data are of very good quality. In Table \ref{T2} we list the nominal elements reported in \cite{BEN} in the framework of a fit which appears to favour small relative inclinations among the planets: a coplanar model is therefore adequate. Semi-major axes and eccentricities are given with very high precision, with the large value $e_1=0.2531$ for the first planet in the resonant chain, planet-c, a Jupiter-like with a period of 30 days. A second gas-giant, planet-b (the first to be discovered) has a period of 61 days and finally there is a Uranus-like object, planet-e with a period of 124.5 days. There is also an internal super-Earth with a period of 2 days not trapped in resonance.

In the context of the present theory, the system exhibits several puzzling aspects which we are going to discuss in what follows. The claim of a system in Laplace resonance is out of question. However, when the configuration is examined, we see that the combination of libration centers is at odds with respect to   the equilibria corresponding to de de Sitter-Sinclair solutions, notwithstanding the unclear evidence for the behaviour of $q_3$. \rm In fact, the third resonant angle apparently rotates according to \cite{RG876} but is found librating around $\pi$  by \cite{MGB}. Then, is reported to be evolving chaotically in \cite{BEN}; however, in this same work the Authors refer also to a chaotic behaviour of $q_4$ which appears to conflict with their own plots. In addition, in the same work, the claim of the description of the chaotic motion of $q_3$ found in \cite{BDH}, is attributed to $q_4$. We remark that, while the outcome concerning $q_3$ appears convincing, it is quite notable to deduce the evolution of the 3-body combination $q_4$ with a 2-planet model. 
  Anyway, the solutions provided by the de Sitter-Sinclair theory predict not only a mismatch in the reported equilibrium value of the resonant angles but, what is more, very different values of the forced eccentricities. However, the quality of the observational data is so good, at least for planet-b and planet-c, that there should be little doubts about the architecture of the system. As can be seen in Table \ref{T2}, the libration centers of $q_1, \ q_2$ and $ q_4$ are found, in the most recent reference \citep{BEN}, all to be zero. On the ground of the 
results obtained here, we can state that the new de Sitter solution denoted as  $X^{(2)}(0)$ provides both the correct architecture and good predictions of the dynamical quantities. \rm 

\begin{table}

    \begin{tabular}{@{}llllcccc@{}}
  \hline\hline
semi-major axis [in unit of $a_1$]                                               		&\rm{Planet c}                   & \rm{Planet b}             	& \rm{Planet e} & & & & \\
\hline

 nominal							                                 		&  1                  			& 1.6074               		&       2.5840   & & & & \\
 de Sitter-Sinclair	                            						&  1                     		& 1.6093                  		&   	 2.6350  & & & & \\
 new de Sitter 		                            						&  1                     		& 1.6052                  		&   	 2.5829  & & & & \\
 \hline\hline
 eccentricity/resonant angles 						& $e_1$		& $e_2$		& $e_3$ 	& $q_1$ \quad & \quad $q_2$& $q_3$ & $q_4$ \\
 \hline
 
nominal	                         									&  0.2531 				& 0.0368 				& 	 0.0310  & 0 \quad & \quad $0$ & rotating? & $0$ \\
de Sitter-Sinclair  ($X^{(1)}(\pi)$)                                  			&  0.0480				& 0.0047				&       0.0064  & $\pi$ \quad & \quad $0$ & $0$ & $\pi$ \\
de Sitter-Sinclair  ($X^{(1)}(0)$)			                                		&  0.0524				& 0.0104				&       0.0023  & $\pi$ \quad & \quad $0$ & $0$ & $0$ \\
new de Sitter ($X^{(2)}(\pi)$)                                  				&  0.2657				& 0.0737				&       0.0117  & 0 \quad & \quad $\pi$ & $\pi$ & $\pi$ \\
new de Sitter ($X^{(2)}(0)$)			                                		&  0.2546				& 0.0366				&       0.0381  & 0 \quad & \quad $0$ & $\pi$ & $0$ \\
new de Sitter ($X^{(3)}(\pi)$)                                  				&  0.2150				& 0.0346				&       0.0095 & $\pi$ \quad & \quad $0$ & $0$ & $\pi$ \\
new de Sitter ($X^{(3)}(0)$)			                                		&  0.2121				& 0.0425				&       0.0094 & $\pi$ \quad & \quad $0$ & $0$ & $0$ \\
\hline\hline
\end{tabular}
\caption{Mean nominal orbital elements of the 3 main planets in GJ-876 and resonant angles according to 
\cite{BEN} compared with predictions from the model.} \label{T2}
\end{table}

  In passing, \rm we observe that the solution denoted as  $X^{(1)}(0)$ is interesting in its own: it is the complementary case with respect to the Galilean solution seen above. It is produced by the condition $\omega>0$ so that the libration center $q_4=0$ turns out to be stable and so the sequence is in this case $q_1 = \pi, \ q_2 = 0 , \ q_3 =0, \ q_4 = 0$. This configuration is considered by \cite{Si75} relevant for the Uranian satellites and has also been discussed by \cite{YoPe} in the context of Galilean dynamics as a possible end-state under dissipative effects. It is however ruled out for GJ-876. 
Rather, the new de Sitter solutions envisaged here, provide quite reliable values for the forced eccentricities: 
  in particular, the solution $X^{(2)}(0)$  is stable and reproduces the observed configuration $q_1 = 0, \ q_2 = 0 , \ q_3 =\pi, \ q_4 = 0$. 
On the other side, the complementary solution $X^{(3)}(0)$ is again stable but has the same configuration  
of $X^{(1)}(0)$ for the Uranian system. 

The second new de Sitter solution $X^{(2)}(0)$ predicts values of the forced eccentricities quite close to the observed ones 
and provides additional convincing dynamical predictions. Triple conjunctions are allowed \citep{RG876}
with planet-c and -b at periastron ($\lambda_1=\lambda_2=\varpi_1=\varpi_2=0$) and planet-e at apastron ($\lambda_3=0,\varpi_3=\pi$). 
By computing the eigenvalues of the stability matrix, we get 
$$
\omega_\mathrm{free} = 0.107, \quad \omega_L=0.011,$$
giving, with the proper units of time, a precession of nodes of $-0.12$ degrees/day 
and a period of the Laplace libration of $2750$ days. These predictions are
quite close to the observed value $-0.11$ degrees/day and $2800$ days reported by \cite{RG876}. \rm

In both cases of 
equilibrium solutions $X^{(2)}(0)$ and $X^{(3)}(0)$, the free eccentricity of planet-e 
  (as large as almost 100\% of the forced one \citep{RG876,BEN}), \rm produces 
a non-librating evolution of $q_3$. For the greatest majority of reasonable initial condition, 
$q_3$ displays a `nodding' behaviour, namely the tendency to repeat patterns of bounded libration 
for several cycles followed by one or more cycles of circulation \citep{KAB}. Since apparently this happens in 
a random way, we may guess a stable chaotic behaviour even if most probably in a small region of phase-space. 
The theoretical and numerical analyses of the 
system have highlighted the chaotic dynamics \citep{MGB,BDH} and diffusion \citep{MCB} in the system. 
However, dynamical stability arguments are invoked to speak of stable chaos 
for which it is reasonable to deduce a dynamical stationary state with well definite, albeit with large amplitude, libration centers. 
We can only remark that these emerge quite clearly in the framework of the model and appear to be fully compatible with the data analysis.  

\section{Conclusions}
\label{SCON}

We have described a comprehensive model for systems locked in the Laplace libration. 
The framework is that of the simplest possible dynamical structure based only on the 
resonant coupling truncated at second order in the eccentricities. The analytic model is then 
constructed by a suitable ordering of the terms in the expansion of the Hamiltonian, 
the study of their equilibria and the computation of resonant normal forms. The agreement 
of the analytic predictions with the numerical integration of the toy model is very good,  

The main result is that of discriminating between two different classes of equilibria, 
according to the sign of the frequency of the free eccentricity oscillations. In the first 
class, only one kind of stable equilibrium is possible: the paradigmatic case is that of the 
Galilean system,   for which a fair reconstruction of the dynamics is achieved when 
including the quadrupole of the Jovian potential by constructing a detuned resonant normal form. 
In the second class, three kinds of stable equilibria are possible and, 
at least one of them, is characterised by a high value of the forced eccentricity for the 
`first planet': here the paradigmatic case is the exo-planetary system GJ-876. 

In the case of the Galilean system, we point out that the resonant normal forms may 
offer useful insights into the evolution of the system under non-conservative perturbations. 
Concerning GJ-876, we are presented with a dynamical configuration with some unexpected 
features (e.g. triple conjunctions of the three planets) which are faithfully reconstructed by 
the new stable equilibria predicted by the model. Here too, the basic Hamiltonian model 
truncated at 2nd order provides a solid basis for the investigation of mechanisms of 
capture to or exit from the resonance. \rm


\section*{Acknowledgements}

This work has been performed with the support of the Italian Space Agency, 
under the ASI Contract n. 2018-25-HH.0 (Scientific Activities for JUICE, C/D phase): 
my deep thanks to the whole Juice teams of Roma Tor Vergata and Roma la Sapienza. I also acknowledge 
fruitful discussions with C. Efthymiopoulos, S. Ferraz-Mello, H. Han{\ss}mann and A. Morbidelli and the
partial support from INFN (Sezione di Roma II) and GNFM-INdAM.

\section*{Compliance with ethical standards}

{  Conflict of interest} The author declares that he has no conflict of interest.

\section*{Appendix: Higher-order normal form for the Galilean system}

In this appendix we provide a preliminary computation of a high-order Laplace 
normal form for the Galilean system. In order to get reliable quantitative 
predictions, the starting model is improved with respect to the toy model of 
subsection \ref{GTM} with the inclusion of: 

\n
1. The oblateness of Jupiter, represented by \citep{La-A}
$$
J_2 = 1.4783 \times 10^{-2}, \quad
R_{\Jupiter}=71398 \; {\rm km}, $$
to be used in the evaluation of the secular precessions as described in 
subsection \ref{DNF}

\n
2. A second-order expansion in the eccentricities of the resonant coupling 
terms as given e.g. in \cite{CePaPu} with Laplace coefficients computed 
for the specific semi-axes ratio corresponding to the de Sitter-Sinclair solution.

\n
With the addition of the contribution due to the oblateness as in \eqref{kappadet}, the semi-slow 
frequencies become
\beqa{GSfreq}
\omega_1 &=& -0.0043, \\ 
\omega_2 &=& -0.0037, \\
\omega_3 &=& -0.0036, 
\eeqa
(tu/T, 1 tu = 1.7714 days) so that the linear part of the Hamiltonian is expressed like that in \eqref{DF41}. 
The corresponding corrections used in the equilibrium solutions of subsection \ref{DSE} lead to 
the eccentricities listed in Table \ref{T1} under the denomination ``de Sitter-Sinclair (analytical with $J_2$)''.

The normal form at order $N$ can be written as a series of the form
\beq{resform}
K^{(N)}= \sum_{n=0}^{N} \sigma^{n} \sum_{{  s},{  k}} a^{(n)}_{{  s},{  k}} {  Q}^{\frac{  s}{2}} e^{i({  k}\cdot{  q})}.
\eeq
At each intermediate order $2 \le n \le N$, the coordinates ${  Q},{  q}$,
with 
$$
{  Q} = (Q_1,Q_2,Q_3,\Lambda), \quad {  q} = (q_1,q_2,q^M_3,\lambda),$$ 
have to be intended as new variables but for ease 
of notation are denoted with the same symbol. The norm of the integer vectors 
$$
{  s} = (s_1,s_2,s_3,s_4), \quad {  k} = (k_1,k_2,k_3,k_4), \quad s_a \in \mathbb N, \quad k_a \in \mathbb Z,$$
at each order must comply with the conditions
$$
|{  s}| = n, \quad 0 \le |{  k}| \le n-1,$$
so that the normal form has the D'Alembert character only with respect to the eccentricity vectors. 
The coefficients $a^{(n)}_{{  s},{  k}}$ are functions of the $\omega_j$ 
and of the parameters $\alpha,\beta_1,\beta_2,\gamma$.

Here we are interested in finding more accurate predictions for the libration frequencies and the resonance width. 
Therefore, we compute the normal form at the Sinclair-de Sitter equilibrium obtaining a series of the form
$$
K_L = \sum_{l, m} a_{l m} \Lambda^{l} \cos m \lambda.$$
In Table \ref{T3} we display its coefficients in a form that can be directly compared with 
the analogous series computed by \cite{He}. The resonance width turns out to be further reduced with respect to 
the value of the toy model and now amounts to
$$
\Delta \Lambda= 0.00019.$$  
The libration periods ($T_j, \, j=1,2,3; T_L$) are listed in Table \ref{T4}: in the second column are 
reported the values computed with the 6th-order normal form \eqref{resform}, showing 
the improvement of the present model with respect to \eqref{GP1}; in the third, are listed the periods  
obtained by averaging the outcomes of numerical integration of the Hamiltonian model described above; 
in the fourth are given the corresponding values obtained by \cite{La-C} from a Fourier analysis of their 
accurate L-1 ephemerides \citep{La-B}.

\begin{table}

    \begin{tabular}{@{}ccccc@{}}
  \hline\hline
                               								& $\Lambda^0 $           			& $ \Lambda^1 $   	& $ \Lambda^2 $ & $ \Lambda^3 $  \\
\hline

-- 							                    		&  --                  					& 0.2170            	&   $- 10.8037$    &  0.2152 \\

 $\cos \lambda$                         						&  $4.8861   \times10^{-5}$      		& $- 0.009808 $        &   	 0.4917  &  -0.1636 \\
  $\cos 2 \lambda$                         					&  $- 3.5413 \times10^{-9}$        	& 0.001150            	&  $- 0.05819$   &  $\dots$ \\
  $\cos 3 \lambda$                         					&  $1.2519 \times10^{-10}$        	&   $\dots$          	&  $\dots$  &  \\
\hline\hline
\end{tabular}
\caption{The coefficients $a_{l m}$ of the libration Hamiltonian normal form.} \label{T3}
\end{table}

\begin{table}

    \begin{tabular}{@{}cccc@{}}
  \hline\hline
 Libration period                              					& Normal form           			& Hamiltonian model   	& \cite{La-C}  	\\
\hline
$T_1$ 							                    	&  412                				& 406            			&   $403.5$     	\\
$T_2$                         							&  479                				& 482            			&   $462.7$  	\\
$T_3$                         							&  492                				& 490            			&   $482.2$  	\\
$T_L$                         							&  2057        					&  2053         			&  2060.0  		\\

\hline\hline
\end{tabular}
\caption{The libration periods in days.} \label{T4}
\end{table}

\bibliographystyle{spbasic}
\bibliography{laplabib}

\end{document}